\documentclass{nature}
\usepackage{epsfig}
\usepackage{amsmath}
\usepackage{amssymb,color}

\begin{document}

\title{Tunable Bistability in Hybrid Bose-Einstein Condensate Optomechanics}

\author{Kashif Ammar Yasir$^{\star}$, Wu-Ming Liu$^{\dag}$}

\maketitle

\begin{affiliations}
\item Beijing National Laboratory for Condensed Matter Physics, Institute
of Physics, Chinese Academy of Sciences, Beijing 100190, China.

$^\star$e-mail: kashif\_ammar@yahoo.com

$^\dag$e-mail: wliu@iphy.ac.cn

\end{affiliations}

\begin{abstract}
Cavity-optomechanics, a rapidly developing area of research, has made a remarkable progress. 
A stunning manifestation of optomechanical phenomena is in exploiting the mechanical effects of light to couple the optical 
degree of freedom with mechanical degree of freedom. In this report, we investigate the controlled bistable dynamics of such 
hybrid optomechanical system composed of cigar-shaped Bose-Einstein condensate (BEC) trapped inside high-finesse optical 
cavity with one moving-end mirror and is driven by a single mode optical field. The numerical results 
provide evidence for controlled optical bistability in optomechanics using transverse optical field which directly interacts 
with atoms causing the coupling of transverse field with momentum side modes, exited by intra-cavity field. This technique 
of transverse field coupling is also used to control bistable dynamics of both moving-end mirror and BEC. The report provides an 
understanding of temporal dynamics of moving-end mirror and BEC with respect to transverse field. Moreover, dependence 
of effective potential of the system on transverse field has also been discussed. To observe this phenomena in laboratory, 
we have suggested a certain set of experimental parameters. These findings provide a platform to investigate the tunable 
behavior of novel phenomenon like electromagnetically induced transparency and entanglement in hybrid systems.
\end{abstract}

In last few years, a lot of investigations have been conducted in the field of cavity-optomechanics. Experimental advances
in cavity-optomechanics have made it possible to couple mechanical resonator with optical degree of freedom\cite{Kippenberg}. 
Another milestone was achieved by the demonstration of optomechanics when other physical objects, most 
notably cold atoms or Bose-Einstein condensate, were trapped inside cavity-optomechanics\cite{Esslinger}. 
Both mirror-field interaction and atom-field interaction are needed to develop numerous sensors and devices 
in quantum metrology, and to study both these phenomenas, we rely on hybrid optomechanical systems. 
In optomechanics, the movable mirror can be cooled down, by the mechanical effects of light, to its quantum mechanical 
ground state \cite{CornnellNat2010,TeufelNat2011,CannNat2011,wmliu1,wmliu2}, thus providing a platform to study 
strong coupling effects in hybrid systems~\cite{Teufel2011,GroeblacherNat2009,ToefelNat2011a,VerhagenNat2012}. 
Such extraordinary investigaions in opto-mechanics motivate researchers in developing gravitational wave detectors~\cite{Braginsky}, 
measuring displacement with large accuracy~\cite{Rugar} and also in developing optomechanical crystals~\cite{EchenfieldNat2012}.
Recent discussion on bistable behaviour of BEC-optomechanical system
\cite{Meystre2010}, high fidelity state transfer \cite{YingPRL2012,SinghPRL2012}, entanglement in optomechanics~\cite{Vitali2012,Vitali2014,Sete2014,Hofer}, macroscopic tunneling of an optomechanical membrane  
\cite{Buchmann2012} and role reversal between matter-wave and quantized light field, are 
directing and facilitating researchers towards achieving new mile-stones in cavity-optomechanics. Furthermore, the 
magnificent work on transparency in optomechanics \cite{agarwal2010,Stefan2010,Peng2014,SafaviNaeini2011}, 
dynamical localization in field of cavity-optomechanics\cite{kashif1,kashif2} and the coupled arrays of micro-cavities\cite{wmliu3,wmliu4,wmliu5,wmliu6} provide clear understanding for
cavity-optomechanics. These notable achievements provide strong foundations to study complex systems 
and cause curiosity among researchers to explore such hybrid 
systems and so, a lot of work has been done in this regard\cite{wmliu7,Yu2012,Kyriienko2014,Jiang2013,hill2012}.

In this report, we investigate tunable bistable behavior of optomechanical system which consists of one fixed mirror,
one moving-end mirror with maximum amplitude $q_0$\cite{Kippenberg} and Bose-Einstein condensate (BEC) trapped inside the cavity, 
driven by a single mode optical field\cite{Zhang2009,Yang2011}. The idea of present work is motivated by the findings of reference\cite{Yang2011}. 
They demonstrated controlled optical switching of 
field inside an optical cavity containing ultra-cold atoms. They used transverse optical field to control bistable 
behavior of intra-cavity photons induced by external pump field. We use this approch of transverse optical field 
to control the bistable behavior of hybrid optomechanical system and show that the optical bistability or photons bistability inside the cavity is increased by
increasing the strength of transverse field. The demonstration of results shows that the transverse optical field can 
also be used to control the bistable dynamics of moving-end mirror as well as BEC. We also investigate how this bistable behavior can affect 
the effective potential of the system. The set of experimental parameters which make these results experimentally
feasible in laboratory has also been provided.

\section{BEC Optomechanics}

Optomechanical system which is a Fabry-P\'{e}rot cavity of length $L$ consists of a fixed mirror and a moving-end mirror driven by a single mode 
parallel pump field with frequency $\omega_{p}$, as shown in Fig.1. Cigar-Shaped BEC, with N-two level atoms, is trapped inside optical cavity
\cite{Esslinger,esteve,Brennecke}. Moreover, optomechanical system is also driven by transverse optical field which interacts 
perpendicularly with BEC with intensity $\eta_\perp$ and frequency $\omega_{\perp}$\cite{Zhang2009,Yang2011,zhang2013}. Counter propagating field inside 
the cavity forms a one-dimensional optical lattice.
The moving-end mirror possesses harmonic vibrations under radiation pressure with frequency $\omega_{m}$, with maximum amplitude $q_0$ and 
exhibits Brownian motion in the absence of intra-cavity field. To make this study of tunable bistability 
in hybrid BEC-Optomechanics experimentally feasible, we suggest a set of particular parameters from present available experimental
setups\cite{Esslinger,Kippenberg,Brennecke,AIP,Kipp07}. We consider $N=2.3\times4$ $^{87}Rb$ atoms trapped inside Fabry-P\'{e}rot cavity with length $L=1.25\times10^{-4}$,
driven by single mode external field with power $P_{in}=0.0164mW$, frequency $\omega_{p}=3.8\times2\pi\times10^{14}Hz$ and
wavelength $\lambda_{p}=780nm$. The frequency of intra-cavity field is 
$\omega_{c}=15.3\times2\pi\times10^{14}Hz$, with decay rate $\kappa=1.3\times2\pi kHz$. Intra-cavity field produces
recoil of $\omega_{r}=3.8\times2\pi kHz$  in atoms trapped inside cavity. The coupling of intra-cavity field with BEC is 
$\xi_{sm}=4.4 MHz$. Detuning of the system is taken as $\Delta=\Delta_{c}+ \frac{U_{0}N}{2}=0.52\times2\pi MHz$, 
where vacuum Rabi frequency of the system is $U_{0}=3.1\times2\pi MHz$. The moving-end mirror of cavity should be a perfect reflector that oscillates with a 
frequency $\omega_{m}=15.2\times2\pi MHz$ and its coupling with intra-cavity field is considered as $\xi=3.8 MHz$.

The complete Hamiltonian of the system consists of three parts,
\begin{equation}
\hat{H}=\hat{H}_{AF}+\hat{H}_{MF}+\hat{H}_{T}\label{1},
\end{equation}
where, $\hat{H}_{AF}$ describes behavior of the atomic mode (BEC) and its coupling with cavity-optomechanics, 
$\hat{H}_{MF}$ is related to the intra-cavity field and its association with the moving-end mirror while, 
$\hat{H}_{T }$ accounts for noises and damping associated with the system. 

We use strong detuning regime to adiabatically eliminate internal excited levels of atomic mode inside the cavity potential. 
We further apply rotating frame at external
field frequency to derive Hamiltonian $H_{AF}$ having quantized motion of atoms along the cavity axis. To avoid atom-atom interactions in BEC, 
we assume BEC is dilute enough so that many-body interaction effects can easily be ignored\cite{Zhang2009,Yang2011}.
\begin{equation}
\hat{H}_{AF}=\int\hat{\psi}^{\dag}(x)\left(-\frac{\hslash d^{2}}{2m_{a}dx^{2}}+
\hslash U_{0}\hat{c}^{\dag}\hat{c}\cos^{2}kx+\eta_{\perp} cos(kx)(\hat{c}^{\dag}+\hat{c})\right)\hat{\psi}(x)dx,  
\end{equation}
$\hat{\psi}(\hat{\psi}^{\dag})$ is bosonic annihilation (creation) operator and 
$U_{0}=g^2_{0}/\Delta_{a}$, $g_{0}$ is the vacuum Rabi frequency and $\Delta_{a}$ is far-off detuning between 
field frequency and atomic transition frequency $\omega_{0}$. The mass of an atom is represented by $m_a$
and $k$ is the wave number for intra-cavity field. $\eta_\perp=g_{0}\Omega_{p}/\Delta_a$ is the coupling of BEC with transverse
field and represents maximum scattering and $\Omega_pp$ is the Rabi frequency of the transverse pump field.
Due to field interaction with BEC, photon recoil takes place that generates symmetric momentum $\pm2{l}\hbar k$ side modes, where, ${l}$ is an integer. 
In the presence of weak field approximation, we consider low photon coupling. Therefore, only lowest order perturbation of the wave function 
will survive and higher order perturbation will be ignored.  
So, $\hat{\psi}$ is defined depending upon these $0^{th}$, $1^{st}$ and $2^{nd}$ modes~\cite{Meystre2010} as,
\begin{equation}
 \hat{\psi}(x)=\frac{1}{\sqrt{L}}[\hat{b_{0}}+\sqrt{2}cos(kx)\hat{b_{1}}+\sqrt{2}cos(2kx)\hat{b_{2}}],
\end{equation}
here, $\hat{b_{0}}$,$\hat{b_{1}}$ and $\hat{b_{2}}$ are annihilation operators for $0^{th}$ 
, $1^{st}$ and $2^{nd}$ modes respectively. By using $\hat{\psi}(x)$ defined above in Hamiltonian $H_a$, we write the Hamiltonian 
governing the field-condensate interaction as,
\begin{eqnarray}
 \hat{H}_{AF}&=&\omega_{r}\hat{b_{1}}^{\dag}\hat{b_{1}}+ 4\omega_{r}\hat{b_{2}}^{\dag}\hat{b_{2}}+
 \frac{U_{0}}{4}\hat{c}^{\dag}\hat{c}[\sqrt{2}(\hat{b_{0}}^{\dag}\hat{b_{2}}+\hat{b_{2}}^{\dag}\hat{b_{0}})
 +2N+\hat{b_{1}}^{\dag}\hat{b_{1}}]  \nonumber \\ 
 &+&\frac{\eta_{\perp}}{2}(\hat{c}^{\dag}+\hat{c})[\sqrt{2}(\hat{b_{0}}^{\dag}\hat{b_{1}}
 +\omega_{r}\hat{b_{1}}^{\dag}\hat{b_{0}})+(\hat{b_{1}}^{\dag}\hat{b_{2}}+\hat{b_{2}}^{\dag}\hat{b_{1}})].
 \end{eqnarray}
The sum of number of particles in all momentum side modes is, 
$\hat{b_{0}}^{\dag}\hat{b_{0}}+\hat{b_{1}}^{\dag}\hat{b_{1}}+\hat{b_{2}}^{\dag}\hat{b_{2}}=N$, where, $N$ is the total number 
of bosonic particles. As population in $0^{th}$ mode is much larger than the population in $1^{st}$ and $2^{nd}$ order side mode, 
therefore, we can comparatively ignore the population in $1^{st}$ and $2^{nd}$ order side mode and can write 
$\hat{b_{0}}^{\dag}\hat{b_{0}}\simeq N$ or $\hat{b_{0}}$ and $\hat{b_{0}}^{\dag}\rightarrow \sqrt{N}$. This is possible 
when side modes are weak enough to be ignored. Under these assumptions, we write simplified form of 
BEC-field Hamiltonian,
\begin{equation}\label{Ha}
\hat{H}_{AF}=\frac{\hbar U_{0}N}{2}\hat{c}^{\dag}\hat{c}+\frac{\hbar\Omega}{2}(\hat{P}^{2}
+\hat{Q}^{2})+\xi_{sm}\hbar\hat{c}^{\dag}\hat{c}\hat{Q}+\hbar\eta_{eff}\hat{Q}(\hat{c}^{\dag}+\hat{c}).
\end{equation}
First term accommodates the effects of condensate on intra-cavity field and $\Delta_{a}$ is the atom-field 
detuning. Second term describes the motion of condensate inside the cavity.
$\hat{P}=\frac{i}{\sqrt{2}}(\hat{b}-\hat{b}^{\dag})$ and 
$\hat{Q}=\frac{1}{\sqrt{2}}(\hat{b}+\hat{b}^{\dag})$ are dimensionless momentum and position operators for 
BEC with commutation relation, $[\hat{Q},\hat{P}]=i$ and $\Omega=4\omega_{r}=2\hbar k^{2}/m_{a}$, is recoil 
frequency of an atom. Third term in eq. (\ref{Ha}) describes coupled energy of field and condensate 
with coupling strength $\xi_{sm}=\frac{\omega_{c}}{L}\sqrt{\hbar/m_{bec}4\omega_{r}}$, where, 
$m_{bec}=\hslash\omega_{c}^{2}/(L^{2}NU^2_{0}\omega_{r})$ is the side mode mass of condensate. The last term accounts
for the coupling of BEC with transverse field and $\eta_{eff}=\sqrt{n}\eta_{\perp}$ is transverse coupling strength.

The Hamiltonian for the part of moving-end mirror $\hat{H}_{MF}$ can be given as\cite{LawPRA1995},
\begin{equation}
\hat{H}_{MF}=\hbar\triangle_{c}\hat{c}^{\dag}\hat{c}+\frac{\hbar\omega_{m}}{2}(\hat{p}^{2}+
\hat{q}^{2})-\xi\hbar\hat{c}^{\dag}\hat{c}\hat{q}-i\hbar\eta(\hat{c}
-\hat{c}^{\dag}),
\end{equation}
first term is associated with energies of the intra-cavity field where, $\Delta_{c}=\omega_{c}-\omega_{p}$ is external 
pump field and intra-cavity field detuning, 
$\omega_{c}$ is intra-cavity field frequency and $\hat{c}^{\dag}$ ($\hat{c}$) are creation (annihilation) operators 
for intra-cavity field with commutation relation  
$[\hat{c},\hat{c}^{\dag}]=1$. Second term describes the motion of moving-end mirror where, $\hat{q}$ and $\hat{p}$ 
are dimensionless position and momentum operators for moving-end mirror respectively, having commutation relation $[\hat{q},\hat{p}]=i$,
which reveals the value of the scaled Planck's constant, $\hbar=1$. 
Third term represents coupling of moving-end mirror with field generated by radiation pressure applied by intra-cavity field on moving-end 
mirror. Here $\xi=\sqrt{2}(\omega_{c}/L)x_{0}$ is the coupling strength and 
$x_{0}=\sqrt{\hbar/2m\omega_{m}}$, is zero point motion of mechanical mirror having mass $m$. 
Last term is associated with the coupling of intra-cavity field with external pump field $\vert\eta\vert=\sqrt{P\kappa/\hbar\omega_{p}}$, 
where, $\kappa$ is cavity decay rate related with outgoing modes and $P$ is the external pump field power.

The Hamiltonian $\hat{H}_{T}$ describes the effects of dissipation and noises associated  
with optomechanical system by using standard quantum noise operators ~\cite{Noise1991}.
The total Hamiltonian $H$ leads to develop coupled quantum 
Langevin equations for optical, mechanical (moving-end mirror) and atomic (BEC) degrees of freedom.
\begin{eqnarray}
\frac{d\hat{c}}{dt}&=&\dot{\hat c}=(i\tilde{\Delta}+i\xi
\hat{q}-i\xi_{sm} \hat Q-\kappa)\hat{c}+i\eta_{eff}\hat{Q}+\eta+\sqrt{2\kappa a_{in}},\label{2a}\\
\frac{d\hat{p}}{dt}&=&\dot{\hat p}=-\omega_{m}\hat{q}-\xi\hat{c}^{\dag}\hat{c}
-\gamma_{m}\hat{p}+\hat{f}_{B},\label{2b}  \\
\frac{d\hat{q}}{dt}&=&\dot{\hat q}=\omega_{m}\hat{p},\label{2c}  \\
\frac{d\hat{P}}{dt}&=&\dot{\hat P}=-4\omega_{r}\hat{Q}-\xi_{sm}\hat{c}^{\dag}\hat{c}
-\gamma_{sm}\hat{P}+\hat{f}_{1m},\label{2d} \\
\frac{d\hat{Q}}{dt}&=&\dot{\hat Q}=4\omega_{r}\hat{P}-\gamma_{sm}\hat{Q}+\hat{f}_{2m}.\label{2e}
\end{eqnarray}
$\tilde{\Delta}=\Delta _{c}-NU_{0}/2$ is the effective detuning of the system and $\hat{a}_{\mathrm{in}}$ is Markovian input 
noise associated with intra-cavity field. The term $\gamma _{m}$ describes mechanical energy decay rate of the moving-end mirror and 
$\hat{f}_{B}$ is Brownian noise operator associated with the motion of moving-end mirror~\cite{Pater06,Giovannetti1}. 
The term $\gamma_{\rm sm}$ represents damping of 
BEC due to harmonic trapping potential which affects momentum side modes while, $\hat{f}_{1M}$ and $\hat{f}_{2M}$ 
are the associated noise operators assumed to be Markovian.

As we focused on the the bistable behavior of optomechanics and its classical dynamics, 
we consider positions and momenta as classical variables. To write the steady- state values 
of the operator we assume optical field decay at its fastest rate so that, we can set time derivative 
to zero in equation (\ref{2a}). The steady-state values of operators are given as, 
\begin{eqnarray}
 c_{s}&=&\frac{\eta+i\eta_{eff}Q}{\kappa +i(\Delta_{a}-\xi q +\xi_{sm} Q)},\\
 q_{s}&=&\frac{\xi c^{\dag}c}{\omega_{m}}\label{equqm},\\
 p_{s}&=& 0,\\
 Q_{s}&=&\frac{-\xi_{sm}c^{\dag}c}{4\omega_{r}[1-\gamma_{a}/4\omega_{r}]}\label{equqa},\\
 P_{s}&=&\frac{\gamma_{a}}{4\omega_{r}}Q_{s}.
\end{eqnarray}\label{equ2}
The steady-state photons number of intra-cavity field is described as,
\begin{equation}
 n_{s}=|c_{s}|^{2}=\frac{\eta^{2}+\eta_{eff}^{2}Q^{2}}{\kappa^{2} + (\Delta_{a}-\xi q +\xi_{sm} Q)^2}\label{equ1}.
\end{equation}

\section{Optical, BEC and moving-end mirror Bistability}

The hybrid BEC-optomechanical system shown in Fig.1 is simultaneously driven by external pump field
with frequency $\omega_p$. When intra-cavity radiation pressure oscillates with 
frequency ($\omega_c$) is equal to the frequency of moving-end mirror $\omega_m$, it generates the Stokes and 
anti-Stokes scatterings of light from the intra-cavity field potential. Conventionally, optomechanical
systems are operated in resolved-sideband regime $\kappa<<\omega_m$, which is off-resonant with Stokes scattering, 
therefore, the Stokes scattering is strongly suppressed and only anti-Stokes scattering will survive inside the cavity 
potential which causes the conversion of external pump photons to the intra-cavity photons. Therefore, by increasing the 
external pump field strength, the anti-Stokes scattering of intra-cavity photon number will be increased and will cause to move 
system in bistable regime. Optical bistability has been discussed in optomechanics in ref.\cite{Jiang2013} by considering 
double cavity configuration and the effects of s-wave scattering on bistable behavior of intra-cavity photon number have been 
studied in ref.\cite{Dalafi}. In our optomechanical configuration, we consider transverse optical field to control the bistable 
behavior of the system. The idea is motivated by the findings of reference\cite{Yang2011}, where, they use transverse optical 
field to control bistable behavior of intra-cavity photons induced by external pump field. This mechanism provides another source 
of inducing photons in the cavity-optomechanics. It can be easily noticed from mathematical description that how this perpendicular channel of photon 
influence the optomechanical system and how it is connected to the displacement of atomic mode of the system. Therefore, the 
transverse photons will scatter inside the cavity depending on the configuration of intra-cavity condensate mode and will behave as 
nonlinear factor to the intra-cavity photon number.

To obtain the steady-state relation for Intra-cavity photon number, we use expression \ref{equ1} and substitute the steady 
state values of position corresponding to moving-end mirror and condensate.
\begin{equation}
 n_{s}=\frac{\eta^{2}+\eta_{eff}^{2}(\frac{-\xi_{sm}n_{s}}{4\omega_{r}[1-\gamma_{a}/4\omega_{r}]})^{2}}{\kappa^{2} 
 + (\Delta_{a}-\xi(\frac{\xi n_{s}}{\omega_{m}}) +\xi_{sm} (\frac{-\xi_{sm}n_{s}}{4\omega_{r}[1-\gamma_{a}/4\omega_{r}]}))^2}\label{equ11}.
\end{equation}
Fig.2 shows the bistable behavior of intra-cavity photon number under the influence of transverse optical field. 
We obtain optical bistability results by solving equ.\ref{equ11}. In Fig.2(a), the blue curve represents
intra-cavity photon number as function of external pump field strength ($\eta/\kappa$) when the transverse field strength
is zero ($\eta_{eff}/\kappa=0$). We can observe, third order roots exist for photon number corresponding to single value
of external pump field, which cause the appearance of bistability in intra-cavity photon number \cite{Yang2011}. 
In Fig.2(a), red curve shows optical bistability
when the value of transverse optical field intensity is $\eta_{eff}/\kappa=200$ and green curve shows optical bistability
when transverse optical field intensity is $\eta_{eff}/\kappa=400$. By observing the red and the green curves in Fig.2(a),
one can easily note that the strength of optical bistability is changed by changing the intensity of transverse optical 
field. The transverse optical field directly interacts with the intra-cavity atomic mode or BEC, 
trapped inside optomechanical system and is the reason for scattering of photons inside the cavity. 
Therefore, optical bistability curve is modified by increasing the strength of perpendicular nonlinearities. In short, photon bistability 
of intra-cavity potential is tunable with the help of perpendicular field. 2(b) shows the continuous 
behavior of intra-cavity photon bistability as a function of external pump field and transverse optical field, which provides
better understanding of the control of intra-cavity optical mode with transverse optical field strength.

Further, to understand the influence of transverse field on intra-cavity optical mode, we study the optical behavior of the system 
with different values of effective detuning. Fig.2(c) shows the intra-cavity optical bistability as a function of external 
pump field and effective detuning of the system in the absence of transverse field. We observe the tilted behavior of photonic 
peaks which represents bistable behavior of photon number. Fig.2(d) and Fig.2(e) represent similar behavior for different 
values of transverse field intensity $\eta_{eff}/\kappa=40, 50$,
respectively. These findings more efficiently clarify the intra-cavity optical mode dependence on perpendicularly interacting 
photons. Initially, intra-cavity optical mode exits only when it is resonant with external pump field. But when we increase the 
strength of external pump field, a tilt appear in the optical mode due to the influence of atomic mode and by increasing the value 
of transverse field, the strength of optical mode show enhancement and it is more tilted towards negative detuning of the system. 
By following these results, we can confidently state that the bistability of optical mode is tunable with the help of transverse 
optical field. 

Transverse optical field cause the the scattering of photons inside the cavity and this scattering can be controlled by 
the strength of transverse optical field. If we increase the strength of transverse field to large values, scattered photons 
inside the cavity will fill the upper branch of bistability and will cause the suppression of optical bistability (see ref.\cite{Yang2011}).
Fig.3(a) demonstrates such behavior of intra-cavity photon at higher value of transverse field. Blue, yellow, red and green curve 
corresponds to the bistable behavior of intra-cavity photon number at transverse field strengths $\eta_{eff}/\kappa=400, 800, 2000, 3000$, 
respectively. We can note that by increasing 
strength of perpendicular field, intra-cavity photons start saturating in upper branch of bistability caused by external driving 
field and the population of 
photons in upper branch becomes much larger than the photons in lower branch of bistablity which leads to the suppression of 
lower stable state. Fig.(b) shows the behavior of upper stable state of bistability with large transverse field strengths. 
Similarly, blue, yellow, red and green curve contain information about the behavior of upper stable state at transverse 
field strengths $\eta_{eff}/\kappa=3000, 3500, 4000, 4500$, respectively and show 
continuous enhancement in the upper branch population with increase in transverse optical field.

To study the bistable dynamics of atomic mode or condensate mode of the system and mechanical mode or moving-end mirror of 
cavity-optomechanics, we rewrite the steady-state solution for position of both modes as,
\begin{eqnarray}
 q_{s}&=&\frac{\xi (\eta^{2}+\eta_{eff}^{2}Q^{2})}{\omega_{m}(\kappa^{2} + (\Delta_{a}-\xi q +\xi_{sm} Q)^2)}\label{equ12},\\
 Q_{s}&=&\frac{-\xi_{sm}(\frac{\eta^{2}+\eta_{eff}^{2}Q^{2}}{\kappa^{2} + (\Delta_{a}-\xi q +\xi_{sm} Q)^2})}{4\omega_{r}[1-\gamma_{a}/4\omega_{r}]}\label{equ13}.
\end{eqnarray}
To discuss bistability effects on the intra-cavity potential, we derive couple equation of motion for both, moving 
end mirror and BEC by treating their motion of moving-end mirror and BEC classically and assuming time scale very short so that
their mechanical damping can be ignored. We write the couple equations of motion by using the quantum Langevin equations as,
\begin{eqnarray}
\frac{d^{2} q}{dt^{2}}&=&-\omega_{m}^{2} q
 +\frac{\omega_{m}\xi(\eta^{2}+\eta_{eff}^{2}Q^{2})}{\kappa^{2}
+(\triangle+\xi q-\xi_{sm} Q)^{2}}\label{3},\\
\frac{d^{2}Q}{dt^{2}}&=&-(4\omega_{r})^{2}Q
-\frac{4\omega_{r}\xi_{sm}(\eta^{2}+\eta_{eff}^{2}Q^{2})}{\kappa^{2}
+(\Delta+\xi q-\xi_{sm}Q)^{2}}\label{3}.
\end{eqnarray}
By using these coupled equation, we derive the effective potential of the intra-cavity field,
\begin{eqnarray}
 V&=&-\frac{\omega_{m}^{2}q^{2}}{2}-\frac{(4\omega_{r})^{2}Q^{2}}{2}+\int\frac{\omega_{m}\xi(\eta^{2}+\eta_{eff}^{2}Q^{2})}{\kappa^{2}
+(\triangle+\xi q-\xi_{sm}Q)^{2}}dq  \nonumber \\ 
&-&\int\frac{4\omega_{r}\xi_{sm}(\eta^{2}+\eta_{eff}^{2}Q^{2})}{\kappa^{2}
+(\triangle+\xi q-\xi_{sm}Q)^{2}}dQ\label{potential}.
\end{eqnarray}
It is known that mechanical degree of freedom (moving-end mirror) and atomic degree of freedom (condensate) are coupled with each 
other through intra-cavity optical mode. The modulation in intra-cavity photon number modify the quantity of radiation pressure 
exerted on atomic and mechanical mode of the system. We have already discussed the bistable behavior of intra-cavity field. Now 
we illustrate that the steady-state position of both atomic mode ($Q_s$) as well as moving-end mirror ($q_s$) possess same 
properties of bistability as optical mode of the system by using equation \ref{equ12} and \ref{equ13}.

Fig.4 demonstrates the bistable dynamics of moving-end mirror of optomechanical system and Cigar-Shaped Bose-Einstein condensate (BEC)
trapped inside optomechanical system in the presence of transverse optical field. Fig.4(a) and Fig.4(b) 
show bistable behavior of moving-end mirror position $q$ and BEC position $Q$
as a function of external field strength in the absence of transverse optical field ($\eta_{eff}/\kappa=0$). 
As equation \ref{equ12} and \ref{equ13} are third-ordered equations for both moving-end mirror and BEC, respectively, so 
by solving these equations, we obtain three roots which lead us to the occurrence of bistable behavior. 
Fig.4(c) shows the effective potential $V$ as a function of mechanical mode and atomic mode steady-state position. 
These results are obtained by solving expression for effective potential \ref{potential} and steady-state values for the position 
of moving-end mirror and BEC. The potential $V$, shown in Fig.4(c), is generally illustrated as two-dimensional double-well potential. 
The two local minima $A$ and $B$ corresponding to the stable points $\{q_{A},Q_{A}\}$ and $\{q_{B},Q_{B}\}$ are shown in 
Fig.4(a) and Fig.4(b). These two points represent two stable regions in effective potential of the optomechanical system 
in accordance to the particular set of values
of moving-end mirror position and BEC position. The point $C$ in effective potential $V$, shown in Fig.4(c) is a 
saddle-point corresponding to values $\{q_{c},Q_{c}\}$ in Fig.4(a) and Fig.4(b). Saddle-point represents unstable behavior of 
the effective potential of system which means at certain particular values of moving-end mirror position and BEC position,
system possesses chaotic behavior, as illustrated in many classical systems such as\cite{Hilborn,Saif2005}.

Fig.4(d) and Fig.4(e) show bistable behavior of moving-end mirror position $q$ and BEC position $Q$ as a function of 
external pump field with transverse optical field strength $\eta_{eff}/\kappa=0.8$ and Fig.4(f) shows their collective 
influence on the behavior of effective potential of optomechanical system. We observe the similar behavior, as we discuss earlier in Fig.4(c),
but due to the presence of transverse optical field, the stable and unstable regions are now modified and the values of stable 
points and saddle-point are changed. The photons interacting perpendicularly with system through transverse field are directly 
associated with intra-cavity atomic momentum side modes which are bounded with mechanical mode (moving-end mirror) in relation 
of radiation pressure. Therefore, a minute variation in the strengths of transverse field cause modified bistable behavior of 
intra-cavity atomic mode $Q_s$ and mechanical mode $q_s$. Moreover, the perpendicularly scattering photons inside the cavity drive 
momentum side modes of BEC and apply radiation pressure on mirror even when there is no external pump field. 
According to equation \ref{potential}, effective potential of system 
completely relies on position of moving-end mirror and BEC. Therefore, a small change in bistable behavior of their positions can bring 
notable modification in effective potential of the system, as shown in Fig.4(f), 
which means the stable points $A$ and $B$, shown in Fig.4(f), are not the same as in Fig.4(c) and saddle-point $C$ is also different from
the saddle-point in Fig.4(c) because of the variations in the bistability of moving-end mirror position and BEC position.
Fig.4(g) and Fig.4(h) illustrate the similar behavior of bistability of moving-end mirror position and BEC position at different 
transverse optical field strength $\eta_{eff}/\kappa=1.6$ and Fig.4(i) shows their mutual effects on the effective potential. 
These results also assert the effectiveness of previous one that the values of both stable points as well as saddle-point are 
changed due to the increase in transverse field strength which cause modification in stable and unstable regions in effective 
potential. By observing these results, one can conclude that the bistable dynamics of moving-end mirror as well as BEC are
tunable with the help of transverse optical field.

It is very important to observe time dependent dynamics of atomic mode and mechanical mode of the system. In this section, 
we will discuss the temporal behavior of $Q$ and $q$ and see how transverse optical field will influence their temporal dynamics. 
Fig.5 accounts for the time dependent response of moving-end mirror and Bose-Einstein condensate corresponding to different 
transverse optical field strengths, in quasi-periodic regime as shown in Fig.4, initially located at $\{q=0,Q=0\}$. 
We solve quantum langevin equation 
to obtain the spatial dynamics of intra-cavity atomic mode 
and mechanical mode of the system under the assumption that the position operator for BEC ($Q$) and moving-end mirror ($q$) behaves like 
classical variables which allow us to ignore quantum noise effects associated with the optomechanical system. 
Fig.5(a) represents
behavior of moving-end mirror with respect to time $\omega_{m}t$ corresponding to different transverse field strengths.
Green line shows the behavior of moving-end mirror in the absence of transverse optical field ($\eta_{eff}/\kappa=0$). The moving-end
mirror follows continuous oscillatory behavior with particular amplitudes corresponding with time as discussed in \cite{Meystre2010}.
The red line represents time dynamics of moving-end mirror when transverse field strength is $\eta_{eff}/\kappa=0.8$. We can clearly 
detect that the amplitudes of oscillations of moving-end mirror have been increased by increasing transverse optical field, 
which cause a shift in moving-end mirror position from stable regime to unstable regime (chaotic regime), as shown in Fig.4(f). 
Similarly, blue curve demonstrates time dependent dynamics of moving-end mirror for $\eta_{eff}/\kappa=1.8$ and shows that 
the oscillatory behavior of moving-end mirror is different from the other curves because of the increase in transverse field.
Transverse optical field directly modulates the intra-cavity field potential which effectively brings nonlinearities to the system 
and therefore, cause chaotic behavior of moving-end mirror. 

Fig.5(b) illustrates time dependent behavior of intra-cavity atomic mode (BEC) under the influence of transverse field strength. 
Green curve is associated with time dependent response of BEC in the absence of transverse field ($\eta_{eff}/\kappa=0$). 
We can note that atomic mode possesses similar temporal oscillatory behavior with particular amplitudes like moving-end mirror. The red line represents time dynamics of BEC at perpendicular optical field ($\eta_{eff}/\kappa=0.8$).
We observe that the amplitude of oscillations of intra-cavity atomic mode is attenuated because
transverse field photons are associated with the position of BEC and cause the suppression of time dependent oscillations 
of atomic momentum side modes.
Similarly, blue curve represents the effects of transverse field ($\eta_{eff}/\kappa=1.8$) on the spatial dynamics of BEC.
Demonstration of these results enable us to conclude that the time dependent dynamics of BEC and moving-end mirror of optomechanics
are also controllable using transverse field.

In order to observe the behavior of effective potential $V$ with respect to the intra-cavity photon number, we derive the 
expression of effective potential (equ. \ref{potential}) by considering the steady-state values for the position of moving-end 
mirror and BEC (equ. \ref{equqm} and \ref{equqa}) as,
\begin{eqnarray}
 V_{s}&=&-\frac{\omega_{m}(\xi n_{s})^{2}}{2}+\frac{4\omega_{r}(\xi_{sm}n_{s})^{2}}{2[1-\gamma_{a}/4\omega_{r}]^{2}}+\int\frac{\xi^{2}(\eta^{2}-\eta_{eff}^{2}(\frac{\xi_{sm}n_{s}}{4\omega_{r}[1-\gamma_{a}/4\omega_{r}]})^{2})}{\kappa^{2}
+(\triangle+\xi (\frac{\xi  n_{s}}{\omega_{m}})+\xi_{sm}(\frac{\xi_{sm} n_{s}}{4\omega_{r}[1-\gamma_{a}/4\omega_{r}]}))^{2}}dn_{s}  \nonumber \\ 
&+&\int\frac{\frac{\xi_{sm}^{2}}{[1-\gamma_{a}/4\omega_{r}]}(\eta^{2}-\eta_{eff}^{2}(\frac{\xi_{sm} n_{s}}{4\omega_{r}[1-\gamma_{a}/4\omega_{r}]})^{2})}{\kappa^{2}
+(\triangle+\xi(\frac{\xi n_{s}}{\omega_{m}})-\xi_{sm}(\frac{\xi_{sm} n_{s}}{4\omega_{r}[1-\gamma_{a}/4\omega_{r}]}))^{2}}dn_{s}\label{potential}.
\end{eqnarray}
Fig.6 demonstrates the relation of effective potential of cavity-optomechanics with the intra-cavity photon number and 
parallel driving field under the influence of transverse optical field. Fig.6(a) shows that effective potential of the 
system increases monotonically with increase in external
pump field and intra-cavity photon number in the absence of transverse optical field. Fig.6(b) represents the similar behavior 
effective potential but in the presence of transverse pump field strength $\eta_{eff}/\kappa=0.8$. Effective potential shows 
monotonic behavior with increase in parallel field but because of the presence of transverse field another curvature appears in 
effective potential. When perpendicular optical field interacts with intra-cavity atomic mode, it causes 
the scattering of photons inside the cavity depending upon the configuration of atomic mode which lead to the modification in 
intra-cavity effective optical potential. We can note the existence of photons inside the cavity in the absence of external pump 
field because of the transverse field scattering from condensates. This scattering also causes the existence of homogeneous atomic side 
modes. 6(b) clearly shows such modulation in the effective potential of the system.

Fig.6(c) represents the response of effective potential inside the cavity at different transverse optical
field strength $\eta_{eff}/\kappa=1.2$. We observe that the curvature appeared in effective potential with intra-cavity 
photon number is enhanced due to the increase in transverse photon scattering. Fig.6(d), (e) and (f) illustrate similar 
behavior of effective intra-cavity potential and photon number with transverse optical field strength $\eta_{eff}/\kappa=1.6$, 
$\eta_{eff}/\kappa=2$ and $\eta_{eff}/\kappa=2.4$, respectively and show continuous enhancement in nonlinear curvature appearing 
in effective potential of the system with transverse photon interaction. These findings lead us to control the collective behavior 
of intra-cavity optical potential with the help of transverse optical field and enable us to understand the explicit nature of 
hybrid optomechanical systems.

\section{Discussion and Conclusion}

In this report, we discuss the bistable dynamics of hybrid BEC-optomechanics. 
We consider $N=2.3\times4$ $^{87}Rb$ atoms trapped inside Fabry-P\'{e}rot cavity with length $L=1.25\times10^{-4}$, with a moving-end mirror and
driven by single mode external field with power $P_{in}=0.0164mW$, frequency $\omega_{p}=3.8\times2\pi\times10^{14}Hz$ and
wavelength $\lambda_{p}=780nm$.  The moving-end mirror of cavity should be a perfect reflector and performs oscillations, because of
intra-cavity field radiation pressure, with frequency $\omega_{m}=15.2\times2\pi MHz$ and its coupling with intra-cavity 
field is considered as $\xi=3.8 MHz$. The frequency of intra-cavity field is 
$\omega_{c}=15.3\times2\pi\times10^{14}Hz$, with cavity decay rate $\kappa=1.3\times2\pi kHz$. Intra-cavity field produces
recoil of $\omega_{r}=3.8\times2\pi kHz$  in atoms trapped inside cavity. The coupling of intra-cavity field with BEC is 
$\xi_{sm}=4.4 MHz$. Detuning of the system is taken as $\Delta=\Delta_{c}+ \frac{U_{0}N}{2}=0.52\times2\pi MHz$, 
where vacuum Rabi frequency of the system is $U_{0}=3.1\times2\pi MHz$. 
We use a transverse optical field, with frequency $\omega_{\perp}$ and coupling $\eta_{\perp}$, to control the bistable dynamics of hybrid system.
Our results demonstrate that the optical bistability of the system is controllable by using transverse optical 
field because the perpendicularly interacting photons are coupled to atomic momentum side modes, exited by the motion of Bose-Einstein condensate due to
the interaction of intra-cavity field. We have also discussed the spatial bistability of moving-end mirror and BEC with different values
of transverse field strength and thus, have shown that the bistable dynamics of moving-end mirror and BEC are tunable using transverse
optical field intensity and studied their effects on effective potential. We further illustrated the temporal behavior of 
mechanical mode and atomic mode (BEC) of the system and found that the dynamics of mechanical mode can be pushed to chaotic regime by 
increasing the strength of vertical field. Besides, we studied the variance in effective potential of the system as a function of 
cavity photon number and external driving field under the influence of transverse field and found out magnificent nonlinear 
response of effective potential with perpendicularly scattering photons. 
This method provides us
an excellent control over the nonlinear dynamics of such hybrid systems.

In future, we will explore the nonlinear dynamics of hybrid system in good cavity limits and will also study the controllable
dynamics by considering interacting Bose-Einstein condensate. Besides, we intend to extend this method to examine the controllability of novel
phenomenon like electromagnetically induced transparency and entanglement. Additional goals include the discussion on the effects of 
spin-orbit coupling using magnetic field in hybrid BEC-optomechanical systems.

\begin{addendum}
\item [Acknowledgement]

This work was supported by the NKBRSFC under grants Nos. 2011CB921502, 2012CB821305, NSFC under grants 
Nos. 61227902, 61378017, 11434015, SKLQOQOD under grants No. KF201403, SPRPCAS under grants No. XDB01020300.
We strongly acknowledge financial support from CAS-TWAS president's fellowship programme (2014).

\item [Author Contributions]
All authors have planned and designed theoretical and numerical studies.
All have contributed in completing the paper.

\item [Competing Interests]
The authors declare that they have no competing financial interests.

\item [Correspondence]
Correspondence should be addressed to Yasir, Kashif Ammar (email:kashif\_ammar@yahoo.com) or to 
Liu, Wu-Ming (email: wliu@iphy.ac.cn).
\end{addendum}

\begin{figure}[h]
\includegraphics[width=7.5cm]{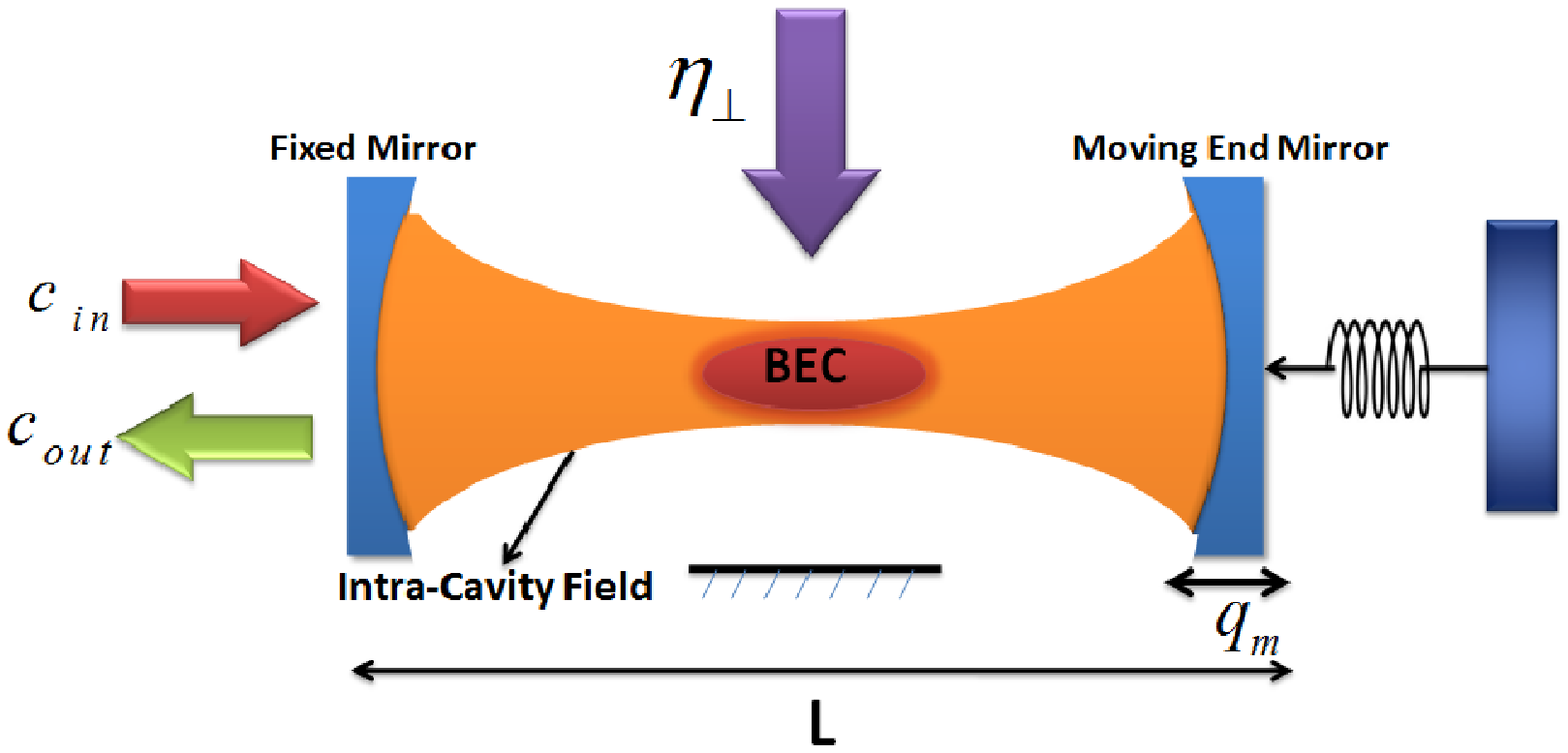}
\end{figure}
\bigskip
\textbf{Figure 1 $\arrowvert$ The hybrid BEC-Optomechanics.}
Cigar-shaped Bose-Einstein condensate is trapped inside a Fabry-P\'{e}rot cavity of length $L$
with a fixed mirror and a moving-end mirror having maximum amplitude of $q_0$ and frequency $\Omega_m$, 
driven by a single mode 
optical field with frequency $\omega_{p}$. Intra-cavity field interacts with BEC and due to photon recoil, the momentum side 
modes are generated in the matter wave. We use another transverse field, with intensity $\eta_\perp$ and frequency $\omega_\perp$, to control the 
bistable dynamics of the system.

\bigskip
\textbf{Figure 2 $\arrowvert$ Tunable optical bistability of intra-cavity field.}
(a) shows optical bistability
as a function of external pump field ($\eta/\kappa$) for different values of transverse optical field. Blue, red and green curves 
are for transverse optical field strength $\eta_{eff}/\kappa=0, 200, 400$, respectively. (b) shows the continuous 
behavior of intra-cavity photon number as a function of interacting transverse field strength ($\eta_{eff}/\kappa$) and external pump field strength ($\eta/\kappa$).
(c) represents optical behavior of intra-cavity field as a function of external field strength ($\eta/\kappa$) and effective detuning ($\Delta/\kappa$) of the system
in the absence of transverse optical field intensity ($\eta_{eff}/\kappa=0$). Similarly, (d) and (e) show
optical bistability as a function of external field ($\eta/\kappa$) and effective detuning ($\Delta/\kappa$) for transverse field strength $\eta_{eff}/\kappa=25, 42$, 
respectively. All numerical parameters are considered from given experimental parameters.

\bigskip
\textbf{Figure 3 $\arrowvert$ Suppression of Bistability with transverse field.}
Suppression of intra-cavity optical mode bistability with large values of transverse optical field. In (a), blue, yellow, 
red and green curve corresponds to the bistable behavior of intra-cavity photon number at transverse field strengths 
$\eta_{eff}/\kappa=400, 800, 2000, 3000$, respectively. (b) accommodates the response of upper stable state of bistability at higher values of 
transverse optical field. Similarly, blue, yellow, 
red and green curve represent upper branch photon number at transverse field strengths 
$\eta_{eff}/\kappa=3000, 3500, 4000, 4500$, respectively. All numerical parameters are used from given experimental parameters.

\bigskip
\textbf{Figure 4 $\arrowvert$ The bistable dynamics of Bose-Einstein condensate and moving-end mirror of cavity.}
The bistable behavior of moving-end mirror position $q$ and Cigar-shaped BEC position $Q$ for the 
different values of transverse field intensities. (a) and (b) show bistability of moving-end mirror position and BEC position, respectively,
as a function of external pump field and in the absence of transverse optical field ($\eta_{eff}/\kappa=0$). (c) demonstrates the effective potential
of the optomechanical systema as a function of moving-end mirror position and BEC position. Points $A$, $B$ and $C$ in (c) correspond 
to the points $\{q_{A},Q_{A}\}$, $\{q_{B},Q_{B}\}$ and $\{q_{C},Q_{C}\}$, respectively, as shown in (a) and (b). 
Similarly, (d) and (e) represent bistability of moving-end mirror and BEC, respectively, at the value of transverse optical field $\eta_{eff}/\kappa=0.8$ and (f) shows
their effects on the effective potential of the system. (g) and (h) represent bistability of moving-end mirror and BEC at
transverse optical field strength $\eta_{eff}/\kappa=1.6$ and (i) shows their effects on the effective potential of the system. 
Similarly, the points $A$, $B$ and $C$, in (f) and (i), correspond to the points given in bistable BEC and moving-end mirror positions results.
All numerical parameters are driven by given experimental parameters. 

\bigskip
\textbf{Figure 5 $\arrowvert$ Temporal behavior of Bose-Einstein condensate and moving-end mirror.} The temporal behavior of 
Bose-Einstein condensate position $Q$ and moving-end mirror $q$ as a function of transverse optical field strength $\eta_{eff}/\kappa$.
(a) describes the time dependent behavior of moving-end mirror position for different transverse field strengths. Green line 
shows moving-end mirror position as a function of time ($\omega_{m}t$) in the absence of transverse optical field ($\eta_{eff}/\kappa=0$). The red line represents
temporal response of moving-end mirror when transverse field strength is $\eta_{eff}/\kappa=0.8$ and blue line
accommodates time dependence of moving-end mirror at transverse field $\eta_{eff}/\kappa=1.8$. (b) demonstrates time dependent
dynamics of Bose-Einstein condensate position corresponding to different transverse field strengths. Green curve corresponds
to the temporal response of BEC in the absence of transverse field ($\eta_{eff}/\kappa=0$). Red and blue lines represent
time dependent behavior of BEC corresponding to $\eta_{eff}/\kappa=0.8$ and $\eta_{eff}/\kappa=1.8$, respectively. 
The strength of external driving field is $\eta=18.4\times2\pi MHz$ and optical detuning of the system is considered as 
$\Delta=0.52\times2\pi MHz$. The remaining numerical 
parameters are same as in previous results.

\bigskip
\textbf{Figure 6 $\arrowvert$ Transverse field effect on effective potential.}
The continuous behavior of effective potential of optomechanical system as a function of intra-cavity photon number 
and external pump field strength with different transverse field strengths. (a) demonstrates the behavior of intra-cavity effective
potential with respect to photon number and external driving field coupling $\eta/\kappa$ in the absence of transverse field $\eta_{eff}/\kappa$.
(b) shows the effective potential when the transverse field strength is $\eta_{eff}/\kappa=0.8$. (c) represents the effects 
of transverse field strength $\eta_{eff}/\kappa=1.2$ on the effective potential. Similarly, (d) shows effective potential for transverse
field $\eta_{eff}/\kappa=1.6$. (e) and (f) demonstrate the effective potential of the system for transverse optical field 
$\eta_{eff}/\kappa=2$ and $\eta_{eff}/\kappa=2.4$, respectively. The remaining parameters are same as in above discussion.

\end{document}